\documentclass{pasj00}
\draft
\begin{document}
\SetRunningHead{Ebisawa, K. et al.}{Spectral Study of the Galactic Ridge X-ray 
Emission with Suzaku}
\Received{//}%{yyyy/mm/dd}
\Accepted{//}%{yyyy/mm/dd}
\title{Spectral Study of the Galactic Ridge X-ray Emission with Suzaku}
%%% begin:list of authors
%
%
 \author{%
   Ken \textsc{Ebisawa},\altaffilmark{1}
   Shigeo \textsc{Yamauchi},\altaffilmark{2}
   Yasuo \textsc{Tanaka},\altaffilmark{3} 
   Katsuji \textsc{Koyama},\altaffilmark{4} \\
   Yuichiro \textsc{Ezoe},\altaffilmark{1}
   Aya \textsc{Bamba},\altaffilmark{1}
   Motohide \textsc{Kokubun},\altaffilmark{1}
   Yoshiaki  \textsc{Hyodo},\altaffilmark{4} \\
   Masahiro \textsc{Tsujimoto}\altaffilmark{5}
   and
   Hiromitsu \textsc{Takahashi} \altaffilmark{6}
%   Greg \textsc{Brown}\altaffilmark{7}
}
 \altaffiltext{1}{Institute of Space and Astronautical Science/JAXA, 
  3-1-1 Yoshinodai, Sagamihara, Kanagawa 229-8510}
 \email{ebisawa@isas.jaxa.jp}
 \altaffiltext{2}{Faculty of Humanities and Social Sciences, Iwate University,
  3-18-34 Ueda, Morioka, Iwate 020-8550}
% \email{yamauchi@iwate-u.ac.jp}
 \altaffiltext{3}{Max-Plank-Instit\"ut f\"ur extraterrestrische Physik,
  D-85740, Garching bei M\"unchen, Germany}
% \email{ki@mpe.mpg.de, ytanaka@xray.mpe.mpg.de}
 \altaffiltext{4}{Department of Physics, Graduate School of Science, 
  Kyoto University, Sakyo-ku, Kyoto 606-8502}
% \email{koyama@cr.scphys.kyoto-u.ac.jp}
 \altaffiltext{5}{Department of Astronomy \& Astrophysics, 
Pennsylvania State University, \\University Park, PA 16802, U.S.A.}
 \altaffiltext{6}{Department of Physical Science, 
Hiroshima University, 1-3-1 Kagamiyama, \\Higashi-Hiroshima, Hiroshima 739-8526}
%%  \altaffiltext{7}{Lawrence Livermore National Laboratory,
%% 7000 East Avenue, L-260, Livermore, CA 94550,  U.S.A.}
%% `\KeyWords{}' always has to be placed before `\maketitle'.
%Do NOT move this preamble from here!

\maketitle

\begin{abstract}
We have observed a  typical Galactic  plane field at $(l,b) = (\timeform{28D46}, \timeform{-0D20})$ with Suzaku\/
for 100 ksec to carry out a precise spectral study of the
Galactic Ridge X-ray Emission (GRXE). The field is known to be devoid of 
X-ray point sources brighter than $\sim2 \times  10^{-13}$ergs s$^{-1}$ cm$^{-2}$ 
(2--10 keV), and already deeply observed with Chandra.
Thanks to the low and stable background and high spectral resolution of Suzaku,
we were able to resolve, for the first time,
 three {\em narrow}\/ iron K-emission lines from low-ionized
(6.41 keV), helium-like (6.67 keV), 
and hydrogenic ions (7.00 keV) in the GRXE spectrum. 
These line features constrain
the GRXE emission mechanisms: The cosmic-ray ion
charge exchange model or the non-equilibrium ionization  plasma 
model are unlikely, since they require either broad emission lines or lines
at intermediate  ionization states.
Collisional ionization equilibrium plasma is the likely origin
for the 6.67 keV and 7.00 keV lines, while origin of the 6.41 keV line,  
which is due to fluorescence from cold material, is not elucidated.
Low non-X-ray background and little stray-light contamination of Suzaku\/ 
allowed us to  precisely
measure the absolute X-ray surface brightness in the direction of the
Galactic plane.
Excluding the point sources brighter than $\sim$2 $\times 10^{-13}$ 
ergs s$^{-1}$ cm$^{-2}$ (2--10 keV), the total surface brightness  on the Galactic 
plane is $\sim$6.1 $\times 10^{-11}$ ergs s$^{-1}$ cm$^{-2}$ deg$^{-2}$ 
(2--10 keV), including the contribution of the cosmic X-ray background that is 
estimated to be $\sim$1.3$\times 10^{-11}$ ergs s$^{-1}$ cm$^{-2}$ deg$^{-2}$. 
% Only $\sim15$ \% of the 2--10 keV GRXE flux is explained by the
% point sources brighter than 
% $\sim$ 3 $\times 10^{-15}$ ergs s$^{-1}$ cm$^{-2}$, which is consistent with 
% the previous Chandra\/ results. 
% Integrating  the known populations of the Galactic point sources down to $\sim10^{-17}$ ergs s$^{-1}$ cm$^{-2}$is not sufficient to account for all the observed GRXE flux.

\end{abstract}

\section{Introduction}
The Galactic plane has been known to be a source of hard (2--10 keV)
X-ray emission since the 1980s (Worrall et al.\ 1982;  Warwick et al.\ 1985; Koyama et 
al.\ 1986). Although most of the X-ray satellites have spent significant amounts of 
observing time to study the Galactic Ridge X-ray Emission (GRXE), its origin has 
not been elucidated yet. Whether the GRXE is truly diffuse emission or composed of 
numerous dim point sources remains  inconclusive. 

Ebisawa et al.\ (2001, 2005) studied a blank Galactic plane area at
$(l,b)\approx(\timeform{28.5D},\timeform{0.0D})$ with Chandra\/, consisting of two partially 
overlapping fields each exposed 100 ksec (sequence 900021 and 900125). They 
found that only 10 -- 15 \%  of the observed 2--10 keV GRXE was resolved 
into point sources brighter than $\sim3 \times 10^{-15}$ ergs s$^{-1}$ cm$^{-2}$.
They also found that the point source surface density in the 2 -- 10 keV band was 
not significantly higher than that at high Galactic latitudes, and therefore 
the majority of these hard X-ray sources is likely to be the background AGNs. 
On the other hand, numerous point sources were detected in the soft ($<2$ keV) band 
which are considered to be nearby Galactic sources, hardly
contributing to the 2 -- 10 keV
GRXE.  Ebisawa et al.\ (2001, 2005) 
suggest that the number of Galactic hard X-ray sources starts to 
deplete ($\log N-\log S$ curve begins to saturate) around 
$\sim 3 \times 10^{-15}$ ergs s$^{-1}$ cm$^{-2}$ (2 -- 10 keV), and they conclude 
that most of the GRXE has the diffuse origin.

Meanwhile, Revnivtsev et al.\ (2006) discovered a striking
similarity between the global GRXE distribution and that of the near infrared (NIR)
emission at 3.5 $\mu$m. Since the NIR emission is considered to represent the 
stellar population, this spatial correlation is suggestive of the stellar origin of 
the GRXE.
Assuming the low-luminosity X-ray source population based on the all-sky RXTE and ROSAT data 
of the sources in the solar neighborhood (Sazonov et al.\ 2006), Revnivtsev et al.\ 
(2006) proposed that the GRXE is ultimately resolved into point sources at 
$\sim10^{-16}$--$10^{-16.5}$ ergs s$^{-1}$ cm$^{-2}$ (2 -- 10 keV). Revnivtsev and 
Sazonov\ (2007) claim that, using the same Chandra\/ data as those of Ebisawa et al.\
(2005), $\sim25 \%$ of the GRXE in 1-- 7 keV
is accounted for by point sources brighter than 
$1.2 \times10^{-15}$ ergs s$^{-1}$ cm$^{-2}$ (interstellar absorption corrected flux), 
and that there is  no indication of saturation of the $\log N-
\log S$ slope down to this flux limit. 
In order to test the point source origin, one would need more than an order 
of magnitude lower limiting flux than the current value. 
%No other than Chandra\/ can do it; yet, it would require much deeper ($\gtrsim$ 1 Msec) exposures.
Currently only Chandra can make the required observations, which would have much deeper
($>$ 1 Msec) exposures than those already carried out.

Another important property of the GRXE is its energy spectrum. It has been known 
from previous observations that the spectrum is hard (e.g, Valinia et al.\ 2000) and 
characterized by many intense emission lines from highly ionized ions of the 
abundant elements (e.g., Kaneda et al.\ 1997).  
These emission lines, in particular the iron K-emission lines in 6--7 keV, can 
constrain the emission mechanism.  
Suzaku\/ satellite is the best suited for this study, since it is equipped with 
the X-ray telescopes of large effective areas, and the focal plane detectors  
of superior spectral resolution. In addition, 
Suzaku\/'s low and stable background and small stray-light contamination enable us
to measure the total surface brightness on the Galactic plane 
accurately, from which we can estimate the absolute GRXE flux by subtracting the
contribution of the cosmic X-ray background. 
In this paper, we report the results of a 100 ksec
observation of the GRXE with Suzaku\/ for the same Galactic plane field as that of 
the Chandra\/ observation by Ebisawa et al (2001, 2005).

\section{Observation and Data Analysis}

\subsection{Observation}
\begin{figure}
\centerline{
\FigureFile(9cm,){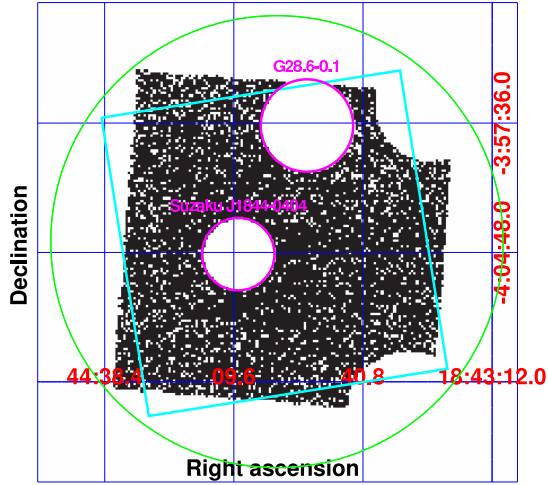}
}
\caption{
The XIS0 sky image of the region selected for the present analysis.
Calibration sources at the upper-right and lower-right corners are removed.
In addition, the supernova remnant
G28.6--0.1 (Bamba et al.\ 2001; Ueno et al.\ 2003; upper-circle in magenta), and the
transient source Suzaku J1844-0404 (Yamauchi et al.\ 2006; lower-circle in magenta) are excluded.
The overlapping square in cyan is the Chandra AO1 (sequence 900021) ACIS field of view.
Assumed sky region to calculate the ARF (green circle with $\timeform{12.'6}$ radius) is also indicated.
}
\label{fig:1}
\end{figure}

%% Suzaku (Mitsuda et al.\ 2007) is the fifth Japanese X-ray satellite launched 
%% on July 10, 2005, which 
Suzaku (Mitsuda et al.\ 2007), the fifth Japanese X-ray satellite, was launched 
on July 10, 2005, and
carries four
X-ray CCD cameras (X-ray Imaging Spectrometers; XIS; Koyama et al.\ 2007a) at the 
foci of 
four X-ray telescopes (XRT; Serlemitsos et al.\ 2007)
and the Hard X-ray Detector (HXD; Takahashi et al.\ 2007).  
The first Suzaku\/ long exposure of the Galactic Ridge (sequence number 500009010) 
was carried out during the proprietary period for the Science Working Group, 
from Oct.\ 28 02:24 to Oct.\ 30 21:30, 2005, for 100 ksec.
The pointing center was chosen exactly at
the same position  as the Chandra AO1 field (sequence number 900021), 
(RA,DEC)=$(\timeform{281.000D}, \timeform{-4.070D})$ (J2000) or 
$(l,b) = (\timeform{28.463D}, \timeform{-0.204D})$. Since the XIS field of view 
($\timeform{17.'8}\times\timeform{17.'8}$) is comparable to
that of Chandra\/ ACIS ($\timeform{16.'9}\times\timeform{16.'9}$), almost the same field was covered 
(figure \ref{fig:1}).

In the present study, we concentrate on the results obtained from the XIS. 
There are four XIS sensors, XIS0 to XIS3, of which XIS1 carries the 
back-illuminated chip and all the rest have front illuminated chips.  
Although XIS1 has superior sensitivity and spectral resolution at low energies 
($\lesssim$ 1 keV), it suffers from a higher intrinsic background relative to 
the front-illuminated sensors. 
The Spaced-row Charge Injection (SCI) operation (Nakajima et al.\ 2007),
which later became a regular operational mode, was not performed in the current 
observation. 
Processing version of the data used in this study is v.1.2.2.3, and the software 
version is heasoft 6.2.

During our observation, a transient source was found at (RA, DEC)=(\timeform{281.034D}, 
\timeform{-4.081D}),
which is named Suzaku J1844--0404 (Yamauchi et al.\ 2007).
To avoid contamination from the source, those events within a two-arcmin radius 
from the transient source are excluded in the following data analysis  
(figure \ref{fig:1}). In addition, we exclude the region which contains a part of 
the supernova remnant G28.6--0.1 (Bamba et 
al.\ 2001; Ueno et al.\ 2003)  
(figure \ref{fig:1}). 
Besides these two, no discrete/point sources brighter than $\sim 2 \times 10^{-13}$ 
ergs s$^{-1}$ cm$^{-2}$ (2 --10 keV) were detected.
In other words, point sources
dimmer than $\sim 2 \times 10^{-13}$ ergs s$^{-1}$ cm$^{-2}$ (2 --10 keV),
either Galactic or extragalactic, are necessarily included in the following GRXE 
spectral analysis.  

\begin{figure*}
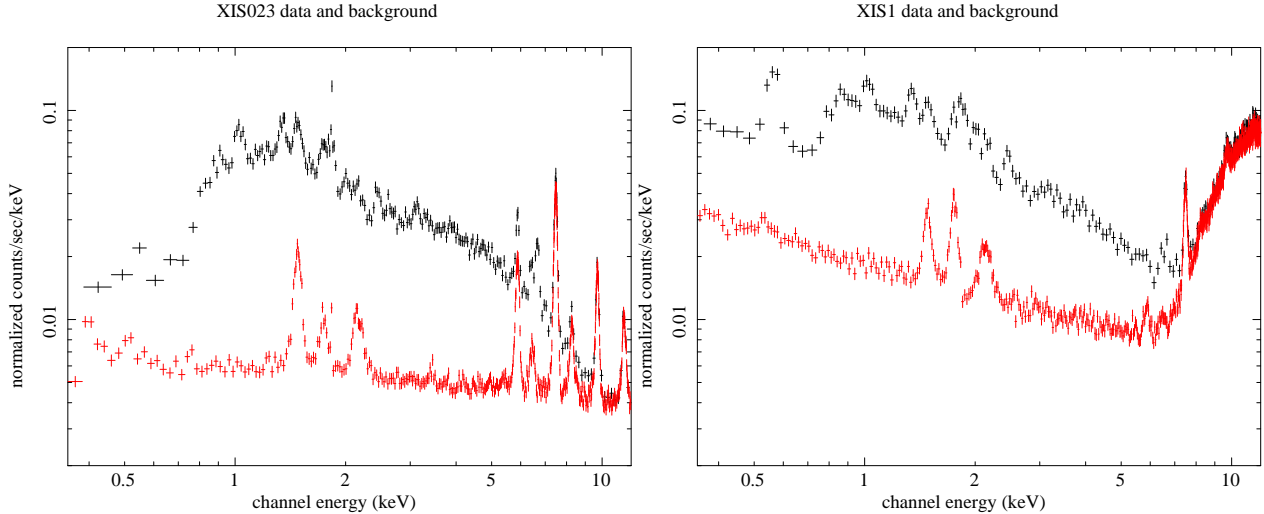

\centerline{
\FigureFile(8.3cm,){figure2-1.eps}\\
\FigureFile(8.3cm,){figure2-2.eps}\\
}
\caption{
The Galactic Ridge spectra (black;  background is not subtracted) and
particle background spectra (red) with XIS0, 2 3 (three sensors averaged; left) 
and XIS1 (right).
}
\label{bgd}
\end{figure*}

\subsection{Background Subtraction}

The intrinsic (non X-ray) background is estimated using 
the XIS background database which was built from the long-term night Earth 
observations 
(Tawa et al.\ 2007)\footnote{http://www.astro.isas.jaxa.jp/suzaku/analysis/xis/nte}. 
For each XIS sensor, the data extraction region is defined by excluding the areas 
of calibration sources and the contamination sources (figure \ref{fig:1}).  
The night Earth background spectra are constructed for the same detector 
region for each sensor.
Since XIS intrinsic background is known to be dependent on the magnetic cut-off
rigidity (COR)(Koyama et al.\ 2007a; Tawa et al.\ 2007), 
the night Earth background spectra were accumulated 
for several different COR bins separately, and properly 
weighted according to the COR distribution during the ridge observation.

Since the characteristics of the three front-illuminated chips are almost identical 
to each other, we combine the data from them and show the average spectrum
(hereafter XIS023).  
In figure \ref{bgd}, we show the observed energy spectra and the corresponding 
night Earth spectra for XIS023 (front-illuminated) and XIS1 (back-illuminated).  

If we compare figure \ref{bgd} with a corresponding (background dominated) figure 
by Chandra (e.g., figure 7 in Ebisawa et al.\ 2005), it is evident that Suzaku\/ 
is much more sensitive to extended emission thanks to the large effective area and 
low background. Among four XIS sensors, XIS0, 2, 3 have much lower intrinsic 
background  in 5 -- 8 keV than XIS1, which is particularly advantageous for the study of the 
iron K-lines.

%% (The difference between the black and red curves in figure \ref{bgd} is the X-ray 
%% emission from the Galactic ridge.  We are not going to subtract the cosmic X-ray 
%% background that is ultimately resolved into many dim AGNs.  Thus, the X-ray 
%% emission we are concerned with in this paper is composed of Galactic point 
%% sources, Galactic diffuse emission, and background AGNs.) *I suggest to delete*

\subsection{Telescope Response}

The telescope response was calculated using the {\tt xissimarfgen}\/ ftool
via ray-tracing (Ishisaki et al.\ 2007). The assumed
sky region of the input photons is a circle with $25.'2$ diameter
(0.139 deg$^2$), that corresponds to  the 
diagonal of a CCD chip (figure \ref{fig:1}).  The uniformly distributed  photons 
from the 0.139 deg$^2$ area
are put into the simulator and
the number of photons falling onto the selected detector area
are calculated for each energy bin. 
Thus, the ``Ancillary Response File'' (ARF) is calculated with the unit of cm$^2$.
Using this ARF in spectral fitting together with the XIS response 
(``Redistribution Matrix File''; RMF)
gives the diffuse spectral model in the unit of photons s$^{-1}$ cm$^{-2}$ keV$^{-1}$
per 0.139 deg$^2$.  In this paper, we give normalized flux in the unit 
of 
photons s$^{-1}$ cm$^{-2}$ deg$^{-2}$.

Suzaku\/ XRT is known to have little stray-light contamination from outside of 
the field of view (\cite{XRT}), in contrast to the similar type telescopes on-board 
ASCA where stray-light significantly affected the GRXE study (e.g., Kaneda et al. 
1997). Consequently, together with the low/stable background, we are able to accurately
measure the energy spectrum and surface brightness of the extended emission
using Suzaku.

\section{Results}

\subsection{Iron K-emission Lines}

\begin{figure}
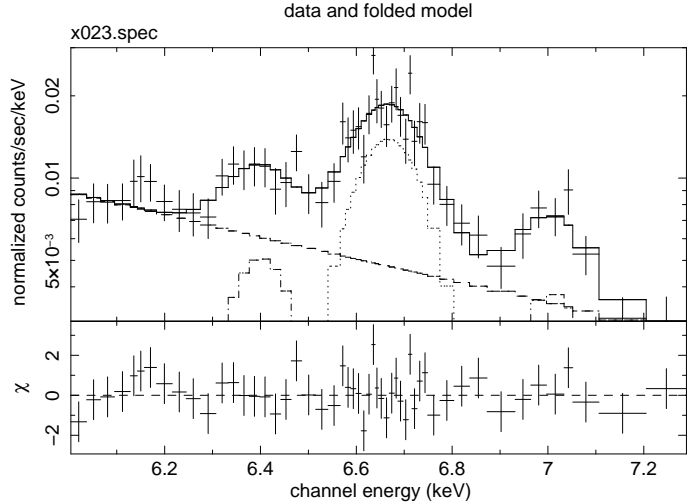

\centerline{
%\FigureFile(8cm,){Ebisawa_Ken_fig1.ps}
\FigureFile(9cm,){ironline.eps}
}
\caption{Spectral fitting of the XIS023 spectrum in the iron K-line region with
a power-law continuum  and three narrow Gaussians.
}
\label{ironline}
\end{figure}

We are particularly interested in the iron K-energy band ($\sim 6.5$ keV), 
since the iron line energies and the widths give us important clues for the GRXE 
emission mechanism. 
We use only the XIS023 spectrum to study iron lines, since XIS1 suffers 
from high background in this energy band (figure \ref{bgd}, right).
The emission line feature is obvious in the XIS023 spectrum. We fitted the data in 
the 6 -- 10 keV band with an absorbed bremsstrahlung model for the continuum 
and three Gaussian profiles with free intrinsic widths
(energies allowed to be free around
6.4 keV, 6.7 keV and 7.0 keV), as well as the neutral $K_\beta$ line
whose energy, intrinsic width (1 $\sigma$) 
and intensity are fixed at  7.06 keV, 10 eV and at 0.125 times the 6.4 keV line 
intensity,
respectively
(Kaastra \& Mewe 1993).
The fit above 6 keV is hardly affected by the hydrogen column density, which
 is fixed at the value constrained from the low energy
model fitting ($N_H = 1.1 \times 10^{21}$ cm$^{-2}$;
 see below). We got $kT$ = 3.9  keV and $\chi^2$=197.6 with d.o.f.=213.
In figure \ref{ironline}, we show the observed energy 
spectrum and the best-fit model.  In table \ref{ironlinetable}, the best-fit line 
parameters are given (errors quoted in this paper correspond to the 90 \% 
confidence limits).
We also tried a 
power-law continuum (photon-index=3.2) instead of the bremsstrahlung, but the 
line parameters hardly  depended on the continuum model used. 

\begin{center}
\begin{longtable}{lccc}
\caption{Iron Line Parameters}\label{ironlinetable}
\hline
\endfirsthead
\endhead
\endlastfoot
Energy [keV]     & $6.41\pm0.02$  & $6.670 \pm0.006$ & $7.00 \pm0.03$\\ % note vol2 p.48
Intrinsic Gaussian width (1$\sigma$) [eV] &     $< 120$        &  $<46$                      & $<52$  \\
Flux [10$^{-5}$ photons/s/cm$^2$/deg$^2$] & $8\pm2$ &  $25\pm3$& $5\pm2$\\
Equivalent Width [eV]  & $80\pm20$ & $350\pm40$ & $70\pm30$ \\\hline
%\end{longtabular}
%\end{center}
\end{longtable}
\end{center}

It is noteworthy that three individual (near neutral, He-like and H-like) iron 
K-lines are clearly visible in the GRXE spectrum for the first time.  
Previously, these lines were only barely resolved with ASCA\/ or Chandra 
because of their poorer 
energy resolutions (e.g., Kaneda et al.\ 1997; Tanaka\ 2002; Ebisawa et al.\ 2005), while
the line fluxes and equivalent widths obtained with ASCA\/ and Chandra\/ 
 are consistent with the present Suzaku\/ values.  

It is also important that we can constrain the intrinsic widths for the 6.67 keV 
line and 7.0 keV line, making reference to the 5.9 keV Mn-K calibration line 
(one sigma width $\approx$ 30 eV; Koyama et al.\ 2007b).
Thus determined widths of 
the iron lines are consistent with 0. %  with the 90\% upper-limit  being 55 eV and 52 eV for 6.67 keV and 7.0 keV line, respectively.

\subsection{Low Energy Lines}

\begin{figure}
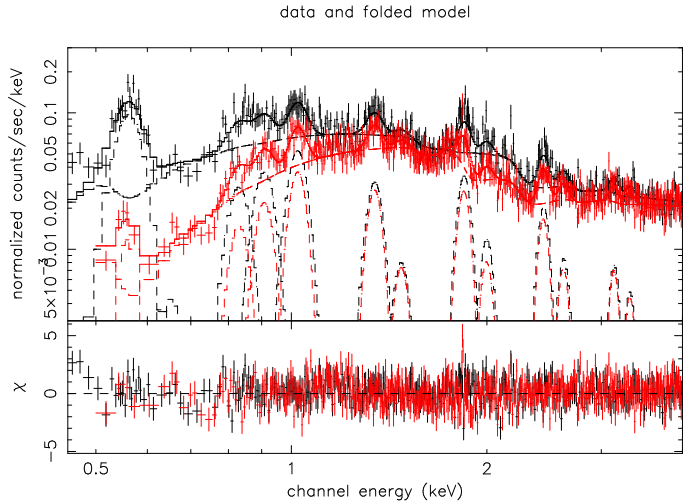

\centerline{
%\FigureFile(8cm,){Ebisawa_Ken_fig1.ps}
\FigureFile(9cm,){lowenergy.eps}
}
\caption{Spectral fitting of the  XIS1 (black) and XIS023 (red) spectra
 in the low energy band.
}
\label{lowenergyline}
\end{figure}

Next, we study low energy lines in the 0.4 -- 4 keV band. Both XIS023 and XIS1 
spectra are fitted simultaneously with the same parameter values except relative 
normalization, where we found XIS1 normalization is 1.03 times that of  XIS023.  
Below, we show normalization values derived from XIS023.  We adopt an absorbed 
power-law for the continuum  and 13 Gaussians  in order to successfully 
reproduce the observed spectrum (figure \ref{lowenergyline}), where we find
choice of the continuum model hardly affects  the line parameters.
The best-fit photon-index and the hydrogen column density are $0.88\pm0.03$ and 
$N_H = (1.1\pm0.1) \times 10^{21}$ cm$^{-2}$,
respectively. Note that these phenomenological values may not represent
actual physical parameters, since origin of the continuum emission is presumably
more complex.
The fit was satisfactory with $\chi^2$ = 1174 for d.o.f=1102.  
Intrinsic line widths are not constrained
besides the strongest O{\footnotesize VII} line (see below), thus we fixed them 
at $1\sigma$= 5 eV.
The best-fit line parameters and the most likely line identifications are given 
in table \ref{lowenergy}. The line flux is not corrected for interstellar 
absorption.

Basically, the same emission lines are detected from highly ionized heavy elements 
as previously observed with ASCA\/ (Kaneda et al.\ 1997; Tanaka\ 2002) and Chandra 
(Ebisawa et al.\ 2005), but with much better resolution. 
Note that the strong O{\footnotesize VII} line at $0.560\pm0.03$ keV is most probably 
dominated by 
the forbidden line (0.561 keV) originated in the Earth's 
magnetosheath caused by solar-wind charge exchange (see 
Fujimoto et al.\ 2007), since the line energy is not consistent with the 
O{\footnotesize VII} resonance line at 0.574 keV that is expected from the GRXE
and/or the local  hot plasma.  Since the 0.560 keV  line is so strong, it
serves for the purpose of calibration;
% The $\sim10$ eV energy shift between the expected energy and the observed energy 
% at the lowest
% energy range is within systematic error of the XIS energy scale (Koyama et al.\ 2007).
the 90 \% upper-limit of the intrinsic line width ($1 \sigma$) is constrained at 6 eV. % (**How is the line resolution calibrated at such low energies? With OVII from SNR?**) 

%% **Remark: the continuum slope is very different between 0.4-4 keV(0.88) and 
%% 6-7.2 keV (4.3). Are we sure that the continuum is abruptly bent? For the ASCA 
%% data, a single power law was OK. If fit over the entire range, say 1-10 keV, do 
%% the iron line result change?**)

\begin{center}
\begin{longtable}{lccl}
\caption{Low energy model fitting  parameters}\label{lowenergy}
\hline
Energy      & Line Flux$^a$ & Equivalent Width & Identification \\
(keV)       & & (eV) &  \\\hline
\endfirsthead
\endhead
$^a$$10^{-5}$ photons s$^{-1}$ cm$^{-2}$ deg$^{-2}$.
\endlastfoot
0.560$\pm$0.003 & $320\pm50$               & $346\pm54$        &O{\footnotesize VII} f, 0.561 keV\\
0.647$\pm$0.024 & 7$^{+13}_{-7}$           & $9^{+17}_{-9}$    &O{\footnotesize VIII}, 0.653 keV\\
0.829$\pm$0.004 & $26\pm7$                 & $40\pm21$         &Fe{\footnotesize XVII}, 0.826 keV\\
0.906$\pm$0.003 & $30\pm7$                 & $49\pm11$         &Ne{\footnotesize IX}, 0.922 keV\\
1.021$\pm$0.002 & $36\pm5$                 & $66\pm9$          &Ne{\footnotesize X}, 1.022 keV\\
1.342$\pm$0.003 & $18\pm3$                 & $42\pm7$          &Mg{\footnotesize XI}, 1.352 keV\\
1.474$\pm$0.012& $4.4\pm2.5 $              & $11\pm6$          &Mg{\footnotesize XII}, 1.472 keV\\
1.843$\pm$0.003 & $22\pm4$                 & $70\pm13$         &Si{\footnotesize XIII}, 1.865 keV\\
1.997$\pm$0.009 & $7.6\pm1.3$              & $25\pm4$          &Si{\footnotesize XIV},  2.006 keV\\
2.444$\pm$0.004 & $21\pm3$                 & $86\pm12$         &S{\footnotesize XV},  2.461 keV\\
2.624$\pm$0.009 & $8.5\pm2.7$              & $36\pm11$         &S{\footnotesize XVI}, 2.622 keV\\
3.139$\pm$0.010 & $7.7\pm2.4$              & $39\pm12$         &Ar{\footnotesize XVII}, 3.140 keV\\
3.321$\pm$0.016 & $5.1\pm2.4$              & $27\pm13$         &Ar{\footnotesize XVIII}, 3.321 keV \\ \hline
\end{longtable}
\end{center}

\subsection{Total X-ray Flux on the Galactic Plane}

As explained above, we successfully fitted the energy spectrum in 6 -- 10 keV and 
0.4 -- 4 keV separately with a simple power-law continuum and Gaussian lines.
However, it is not easy to find a model which can fit the observed spectrum in 
the entire energy range.  
Therefore, in order to determine the total flux in the direction of  the Galactic plane 
in the standard energy band  2 -- 10 keV, we furthermore
fit the spectrum in 4 -- 6 keV separately with simple power-law, 
and sum the contributions from the individual energy bands. 
%% (**Question: what was the slope in the range 7.2-10keV? It cannot be very steep 
%% like 4.3. Otherwise it contradict with previous result, e.g. Valinia et al.**)

The 2 -- 10 keV flux thus determined is 
$6.1 \times 10^{-11}$ ergs s$^{-1}$ cm$^{-2}$ deg$^{-2}$ 
(XIS0, 2 and 3 average).
Also, the 0.5 -- 2 keV flux obtained from  the low energy spectral fitting 
(previous section)
%is $1.41 \times 10^{-11}$ ergs s$^{-1}$ cm$^{-2}$ deg$^{-2}$ (XIS0, 2 and 3 average).
%(**comment: give the value excluding the OVII line here**)
is $1.1 \times 10^{-11}$ ergs s$^{-1}$ cm$^{-2}$ deg$^{-2}$ (XIS0, 2 and 3 average),
excluding the O{\footnotesize VII} forbidden line.
Note that these fluxes include contribution of point sources 
dimmer than $\sim 2 \times 10^{-13}$ ergs s$^{-1}$ cm$^{-2}$ (2 -- 10 keV), which 
are not resolved with Suzaku\/.

The Chandra\/ fluxes calculated with the same condition are 
$7.4 \times 10^{-11}$ ergs s$^{-1}$ cm$^{-2}$ deg$^{-2}$ (2 -- 10 keV) and
$9.7 \times 10^{-12}$ ergs s$^{-1}$ cm$^{-2}$ deg$^{-2}$ (0.5 -- 2 keV) (Ebisawa et al.\ 2005). 
%The $\sim$20 \% difference in the 2 -- 10 keV flux is considered to be within the systematic uncertainty of calibration between Chandra\/ and Suzaku\/.  
The $\sim$20 \% difference in the 2 -- 10 keV flux is due to 
systematic calibration uncertainty between Chandra\/ and Suzaku, which has yet to be understood.
%On the other hand, the significant excess
% of the Suzaku flux in 0.5--2 keV is primarily due to 
%contribution from the 0.574 keV line, which was not 
%taken into account in the Chandra flux calculation since it was 
%not clearly seen.

%% The above value still includes the contribution of the cosmic X-ray background 
%% (CXB). As described in detail in the next section, we estimate the CXB contribution 
%% based on the composite  model of Moretti et al.\ (2003) and applying the interstellar 
%% absorption. Assuming an absorption column of $\sim 6 \times 10^{22}$ 
%% H atoms cm$^{-2}$ through the Galactic plane (see Ebisawa et al.\ 2001), 
%% $\sim1.3 \times 10^{-11}$ ergs s$^{-1}$ cm$^{-2}$ deg$^{-2}$ is expected from 
%% the CXB behind the plane. Although this value is subject to a considerable 
%% fluctuation caused by a few relatively bright AGNs, it is still a minor fraction 
%% of the observed total surface brightness. 
%% Thus, we conclude that the present observation gives the GRXE flux (only Galactic
%% components included) of 
%% $\sim 4.8 \times 10^{-11}$ ergs s$^{-1}$ cm$^{-2}$ deg$^{-2}$. 
%% % (**Exercise: derive the volume emissivity **)

% \subsection{Background Subtraction}

%% \section{Discussion}

%% \subsection{Origin of the Iron Emission Lines}

%% \subsection{Contibution of Point Sources}

\section{Discussion}

\subsection{Emission Mechanism of the GRXE}

We have observed a typical Galactic plane region at 
$(l,b)\approx(\timeform{28.46D},\timeform{-0.20D})$ (= the Chandra\/ AO1 deep field) with Suzaku\/.  
Thanks to the superior spectral resolution, the large effective area and the 
low background of Suzaku\/ XIS, we obtained so far the best quality spectrum of 
the GRXE. For the first time, the iron K-line complex was clearly resolved into
three individual peaks at 6.4 keV, 6.67 keV and 7.0 keV. Importantly, 
the line widths are also tightly constrained to be less than $\sim$50 eV (1 $\sigma$).  

The central energy of the He-like K$\alpha$ line, which   is in fact 
a blend of the resonance, inter-combination and
forbidden lines,  is a good indicator of the plasma condition
 (see Koyama et al. 2007b). In the case of collisional ionization 
equilibrium (CIE), the centroid is 6.680--6.685 keV, depending on 
 plasma densities and different 
plasma models (Koyama et al. 2007b).  The peak energy  is shifted downward
if the  plasma is in a  non-equilibrium ionization  (NEI) state.  
Using ASCA SIS, Kaneda et al.\ (1997) 
claimed to have detected a {\em single}\/ narrow 
emission line at $6.61\pm0.04$ keV, that is significantly lower than what is expected in the CIE case, and suggested that the iron line is from 
a NEI plasma.
Chandra observation by Ebisawa et al.\ (2005)  also indicated a similar 
line center energy $6.52\pm^{0.08}_{0.14}$ keV, if the line feature is
fitted with a {\em single}\/ Gaussian.
Meanwhile, thanks to Suzaku's superior sensitivity and 
energy resolution, we unambiguously detected the two separate lines
at 6.41 keV and 6.670 keV, the flux of the former being
one-third of that of the latter.
The centroid of the two  lines will be at 6.58 keV, and the
7.0 keV line is too weak to be  detected with ASCA and Chandra.
Hence, we consider that
the
lower He-like K$\alpha$  line energies previously reported by  ASCA and Chandra were significantly 
 affected by the
presence of the 6.41 keV line,  which was unresolved by these instruments.

Koyama et al.\ (2007b) determined the He-like K$\alpha$ line energy from the 
Galactic Center (GC) plasma as $6.680\pm0.001$ keV, which is consistent with the CIE plasma model.
We obtained a very close value of  $6.670\pm0.006$ keV, which
does not show evidence of  the NEI conditions.
Thanks to the brightness of the GC plasma illuminating  the entire CCD chip, 
Koyama et al.\ (2007b) were able to perform the chip-segment dependent Charge Transfer
Inefficiency (CTI) correction, which
effectively increased the iron line energies by several electron-volts.
We cannot apply the same correction,  since the Ridge emission is not so strong;
however,  if we consider the possible upward shift of the line energies due to 
this effect,
our line energy completely agrees with that from the GC and the CIE plasma.
The observed  H-like/He-like line intensity 
ratio ($\sim 0.2$) is also consistent with that expected 
for a CIE plasma of $kT$ = 5 -- 8 keV. 

Earlier, Tanaka et al. (2000) reported that, analyzing the ASCA\/ data of the 
extended emission in the  Galactic center region, the iron K-lines are 
significantly broadened with a width of $\sim 70$ eV. They suggested a possibility 
of the charge exchange process for the origin of the lines, where low-energy 
cosmic-ray iron ions will undergo charge exchange as they slow down in the interstellar space.  
They interpreted the finite line width as due to the bulk motion of the cosmic-ray ions. 
Tanaka (2002) pointed out the close similarity of the shape of the GRXE spectrum 
(at the same direction as the present one) with that around the Galactic center, 
and considered that the iron lines in the GRXE may as well be broadened, while
this could not be  examined with ASCA due to insufficient statistics.

The He-like K$\alpha$ line energy expected from the charge exchange process is
$6.666\pm0.005$ keV (Wargelin et al.\ 2005), which is not clearly excluded 
from the present result $6.670\pm0.006$ keV unless the segment-dependent CTI correction is applied.
However, our result supports that 
the He-like and H-like iron lines from the Galactic ridge are essentially 
narrow lines, which is against the line-broadening due to 
the cosmic-ray bulk motion and charge exchange.
The Suzaku\/ observation of the GC region also clearly refuted the broadening 
of the iron lines, as well as that the He-like K$\alpha$ line energy does not agree with that
expected from the charge exchange 
 (Koyama et al. 2007b).  % (**comment: this is conclusive if the line width upper limit is 30 eV. 50 eV will make the argument controversial.**) 
Therefore, we conclude that the charge exchange origin is quite unlikely for 
GRXE,  as well as GC plasma.\footnote{The reason why Tanaka et al. (2000) obtained the line width of $\sim70$ 
eV is still unclear. As written in their paper, the observed spectra from eight 
CCD chips were stacked with reference to the 6.7-keV line peak position, such that 
the line width after summation was minimized. One possibility is that the tuning 
of the energy scales among each other was not sufficient.}

%These two models are unlikely, since they are not consistent with the presence of
%three narrow emission lines. In particular,  
%charge exchange reaction between the cosmic-ray iron ions and H or H$_2$ is most 
%frequent
%at the iron ion velocity of $\sim$5000 km s$^{-1}$ (Tanaka et al.\ 2000, Tanaka 
%2002).
%On the other hand, current upper-limit of the intrinsic width ($\sim$50 eV) of 
%the 6.7 keV line
%or 7.0 keV line corresponds to $\sim2000 $ km s$^{-1}$, thus the 
%charge exchange reaction is unlikely.  

Thus, a CIE plasma is the most probable origin of the He-like and H-like
iron K$\alpha$ lines. By the same token, 
the lower-energy lines from highly ionized ions of other abundant elements are 
considered to be thermal emission. However, the thermal structure of the plasma 
is not simple. As shown in Table 2, the H-like/He-like line intensity ratio is 
similar among different elements from Mg through Ar, and all roughly 0.3 -- 0.5,
and the same is true for Fe. This presumably requires a 
multi-temperature plasma. 

There exist two alternative possibilities to explain the CIE plasma emission from 
the Galactic Ridge. 
The fundamental difference is whether such plasma is genuinely diffuse or 
gravitationally bound in the stellar sources. 
Consequently the ultimate questions are the following: 
(1) Does the superposition of all the point sources reproduce the observed GRXE 
spectrum, in particular, the complex emission line features?
(2) Is the number of such Galactic point sources sufficient to explain the total 
GRXE  flux?

As regards the origin of the 6.4 keV line, which is not
emitted  by the hot plasma but due to fluorescence from cold material,
 another explanation is required. Based on the diffuse model, 
Valinia et al.\ (2000)  proposed that
interaction between the interstellar medium  and cosmic supra/non-thermal electrons 
is responsible for the 6.4 keV line, as well as for the hard-tail observed
above $\sim10 $ keV (Yamasaki et al.\ 1997; Valinia et al.\ 2000).
On the other hand, superposition of numerous, different kinds of  
point sources may create such complex iron line spectra (Revnivtsev et al.\ 2006). In fact,
quiescent cataclysmic variables are known to emit three iron lines, where the 6.4 keV line
is due to reflection from the white-dwarf surfaces (e.g., Ezuka \& Ishida 1999). 
Also, cataclysmic variables often 
reveal multi-temperature spectra with various emission lines as seen in GRXE.

%% \begin{figure}
%% \centerline{
%% %\FigureFile(7.5cm,){Ebisawa_Ken_fig2.ps}
%% \FigureFile(8.6cm,){cumulflux3.eps}
%% }
%% \caption{Total surface brightness on the Galactic plane  2 -- 10 keV measured with 
%% Suzaku\/ 
%% and Chandra\/ (horizontal lines in red and black, respectively)
%%  excluding the point sources brighter than 2 $\times 10^{-13} $
%% ergs s$^{-1}$ cm$^{-2}$.  Cumulative flux of the 
%% Chandra point sources, both Galactic and extragalactic, brighter than 3 $\times 
%% 10^{-15} $
%% ergs s$^{-1}$ cm$^{-2}$  in the same field is  indicated in black.
%% Expected cosmic X-ray background flux through the Galactic plane
%% (blue horizontal line), as well as cumulative 
%% extragalactic point source fluxes (shaded area in cyan), are also indicated 
%% for comparison (Moretti et al.\ 2003).  Orange hatch area indicates the 
%% model of cumulative Galactic source fluxes calculated  by Revnivtsev et al. (2006)
%% for $l=\timeform{20D}$.
%% }
%% \label{grxeflux}
%% \end{figure}

\subsection{Absolute flux of the GRXE}

Excluding the point sources brighter than $2 \times 10^{-13}$ ergs s$^{-1}$cm$^{-2}$
(2-- 10 keV), we have measured the total flux in  the direction of the 
Galactic plane (including cosmic X-ray background)
as  $6.1 \times 10^{-11}$ ergs s$^{-1}$ cm$^{-2}$ deg$^{-2}$ (2--10 keV), which is
consistent with the Chandra\/ measurement on the same region within 20 \%.
%These are shown in figure \ref{grxeflux} as red and black horizontal lines, respectively. 
%Also in figure \ref{grxeflux}, the black curve is the 
% cumulative flux of the  Chandra\/  point sources (both Galactic and extragalactic)
%in the same field brighter than $3 \times 10^{-15}$ ergs s$^{-1}$cm$^{-2}$ (Ebisawa et al.\ 2005).
%If the GRXE is ultimately resolved into point sources (e.g., Revnivtsev et al.\ 2006),
%extrapolation of the black curve should eventually
%reach the upper horizontal lines, but 
Currently, 
only $\sim$15 \% of the 2 -- 10 keV of the total surface brightness is explained by 
the  point sources (either Galactic or extragalactic) brighter than  $\sim3 \times 10^{-15}$ ergs s$^{-1}$cm$^{-2}$  (Ebisawa et al.\ 2005).
In 1 -- 7 keV, where Galactic point sources are more dominant, $\sim25$ \% of the total flux 
is resolved into point sources with interstellar absorption corrected fluxes
higher than $ 1.2 \times 10^{-15}$ ergs s$^{-1}$cm$^{-2}$ 
(Revnivtsev \& Sazonov 2007).

For comparison, we estimate  the cosmic X-ray background (CXB) flux seen through the
Galactic plane.  
The 2-- 10 keV CXB flux determined from the high galactic
region is $\sim2 \times 10^{-11}$ ergs s$^{-1}$ cm$^{-2}$ deg$^{-2}$ (Moretti et al.\ 2003),
where point sources dimmer than $8.0 \times 10^{-12}$ ergs s$^{-1}$ cm$^{-2}$ are included.
In order to compare with our results where  point sources brighter than
$2.0 \times 10^{-13}$ ergs s$^{-1}$ cm$^{-2}$ are excluded, we subtract the cumulative flux of the
sources between 
 $2.0 \times 10^{-13}$ ergs s$^{-1}$ cm$^{-2}$ and $8.0 \times 10^{-12}$ ergs s$^{-1}$ cm$^{-2}$  
using the model extragalactic $\log N - \log S$
curve by Moretti et al.\ (2003); thus we obtained the unabsorbed CXB flux
$\sim1.8 \times 10^{-11}$ ergs s$^{-1}$ cm$^{-2}$ deg$^{-2}$.

The hydrogen column density through the 
Galactic plane is estimated to be
 $\sim 6 \times 10^{22}$  cm$^{-2}$ in this direction (Ebisawa et al.\ 2001), which attenuates the
background CXB down to $\sim70$ \% in 2 -- 10 keV, assuming a  photon-index of 1.7.
Consequently,  $\sim1.3 \times 10^{-11}$ ergs s$^{-1}$ cm$^{-2}$ deg$^{-2}$ is expected from the
background CXB on the Galactic plane, which is ultimately resolved into individual AGNs.
%, which  is indicated as the horizontal blue line in figure \ref{grxeflux}.
%100 \% of the CXB is presumably  resolved into AGNs.
%Moretti et al.\ (2003) of the extragalactic sources 
% is also indicated in blue on figure \ref{grxeflux}. 
Chandra\/ already resolved point-sources down to $3 \times 10^{-15}$ ergs s$^{-1}$cm$^{-2}$, 
whose cumulative flux is $\sim9 \times 10^{-12}$ ergs s$^{-1}$ cm$^{-2}$ deg$^{-2}$.
Therefore, we see that a significant part of the dim hard
Chandra\/ sources is extragalactic.  In fact, only $\sim$20 \% of the
Chandra\/ hard X-ray sources had near-infrared counterparts down to $K\approx16$ (hence considered to
be Galactic), 
whereas almost all the soft X-ray sources had counterparts (Ebisawa et al.\ 2005).

%% Finally, we compare our results with the Galactic point source model
%%  by Revnivtsev et al.\ (2006).  The model cumulative flux of the Galactic sources
%% (figure 9 in Revnivtsev et al.\ 2006)
%% is shown with the orange hatched area in figure \ref{grxeflux}.  
%% Compared to our measurement (black curve), the
%% orange hatch obviously {\em overestimates}\/ the cumulative 
%% Galactic source fluxes.  There are presumably  two  reasons for this:
%% The first reason is that the cumulative model flux by
%%  Revnivtsev et al.\ (2006) includes the point sources as bright as 
%% $3 \times 10^{-12}$ ergs s$^{-1}$ cm$^{-2}$, while we exclude sources brighter than 
%% $2 \times 10^{-13}$ ergs s$^{-1}$ cm$^{-2}$.  The second reason is that the model is made for
%% $l=\timeform{20.0D}$, where point sources are more populous than our observation at $l=
%% \timeform{28.5D}$.
%% Therefore, we need to normalize the orange curve to match our observation.
%% In fact, observed point source fluxes  (black curve) are 
%%  composite of the extragalactic sources
%%  (blue curve) {\em and}\/ the
%%   Galactic sources (orange curve).  
%% In order to make sum of the extragalactic source flux  and  the Galactic source flux
%% be consistent with the Chandra cumulative point source flux, we need
%% to shift the orange curve downward by a factor of 2 to 3.

In conclusion, we measure the flux of the the Galactic Ridge X-ray emission 
as $\sim4.8\times 10^{-11}$ ergs s$^{-1}$ cm$^{-2}$ deg$^{-2}$ (2 -- 10 keV), 
excluding the Galactic and extragalactic sources brighter than $2 \times 10^{-13}$ ergs s$^{-1}$cm$^{-2}$ and the estimated contribution of  dimmer extragalactic sources.
How much fraction of this emission is ultimately resolved into Galactic point sources
has been a long standing question (e.g., Worrall \& Marshall 1983; 
Ottmann \& Schmitt 1992; Mukai \& Shiokawa 1992; 
Ebisawa et al.\ 2005; Revnivtsev et al.\ 2006; Revnivtsev \&
Sazonov 2007).  A future ultra-deep Chandra  observation on the Galactic Ridge
is expected to give the answer.

%% even if we integrate all the Galactic point sources down to
%% $\sim10^{-17}$ ergs s$^{-1}$ cm$^{-2}$ (2 -- 10 keV), the total Galactic cumulative
%% source flux will be $(1 - 3) \times 10^{-11}$ ergs s$^{-1}$ cm$^{-2}$ deg$^{-2}$,
%% as opposed to the observed GRXE flux   $\sim4.8 \times 10^{-11}$ ergs s$^{-1}$ cm$^{-2}$ deg$^{-2}$
%% (excluding the CXB).
%% Hence, all the observed GRXE flux is {\em not}\/
%%  explained by the known point sources brighter than $\sim10^{-17}$ ergs s$^{-1}$ cm$^{-2}$ (2 -- 10 keV).
%% Integrating further dimmer sources would not help significantly, 
%% since such dim sources may contribute in  number,
%% but not contribute to  the cumulative flux.

\subsection{Plasma Density Diagnostics of the GRXE}

We obtained the GRXE flux (excluding the extragalactic sources)
as  $\sim4.8\times 10^{-11}$ ergs s$^{-1}$ cm$^{-2}$ deg$^{-2}$ (2 -- 10 keV)
in the direction of $(l,b)\approx(\timeform{28.46D},\timeform{-0.20D})$.
Assuming that distance to the Galactic center is  8.5 kpc and radius of the
Galactic disk 20 kpc, distance to the edge of the Galactic disk in this
direction  is $\sim25$ kpc. Hence, we estimate the GRXE
emissivity as $\sim2.6\times10^{-29}$ erg s$^{-1}$ cm$^{-3}$ or $\sim7.7 \times 10^{26}$
erg s$^{-1}$ pc$^{-3}$.
Thermal bremsstrahlung emissivity is given as $\sim2.4 \times 10^{-27}\: T^{1/2}\:  N_e^2 
$ erg cm$^{-3}$ s$^{-1}$ (Zombeck 2007), thus if we assume uniformly distributed
hot plasma with $T \approx 6
\times 10^7 $K,  the electron density is $N_e \sim 1.2 \times 10^{-3}$ cm$^{-3}$,
which is extremely tenuous.

The Galactic volume corresponding to 1 deg$^2$ in the direction of
$(l,b)\approx(\timeform{28.5D},\timeform{0.0D})$ is $\sim1.6 \times 10^9$ pc$^3$.
If the GRXE is ultimately resolved into point sources such as
cataclysmic variables or active binary stars, 
$\sim10^5$ sources are expected per
deg$^2$ (Revnivtsev et al.\ 2006).  Hence the spatial density of such sources is
 $\sim6.3 \times 10^{-5}$ pc$^{-3}$, and the average luminosity is $\sim1.1 \times 10^{31}$ erg s$^{-1}$
per source.  
We estimate volume of the plasma associated with each point source as $(f  R_\odot)^3$,
where $f$ is the scaling factor, $\approx 1$ for active binaries and $\lesssim 0.01$
for cataclysmic variables (white dwarfs). In the case of magnetic cataclysmic variables,
which is supposed to be responsible for hard ($\gtrsim$10 keV) part of the GRXE
 (Revnivtsev et al.\ 2006),  
$f$ is expected to be
particularly small, since accretion flow is concentrated on the small magnetic pole regions.
Again assuming $T \approx 6 \times 10^7 $K, we obtain $N_e \sim 4 \times 10^{10} \: f^{-3/2}$ cm$^{-3}$.

Thus, we see that the GRXE plasma densities expected for the diffuse case and the point source
case are  many orders of magnitude different, which is distinguishable from plasma diagnostics. 
In near future, if we could resolve the He-like K$\alpha$ complex into
individual lines, we may constrain density of the plasma, hence
origin of the plasma emission.
As a matter of fact, the resonance line (w) at 6699 eV is always the strongest, whereas
emissivity of 
the forbidden line (z) at 6634 eV and those of the inter-combination lines (x,y)
at 6680 eV and 6665 eV are dependent on the plasma density.
In the high density plasma,
the forbidden line is not  supposed to be observed, while in the low-density 
coronal limit, the forbidden line is stronger than the inter-combination lines.
Such precise spectral observations with a few eV  resolution  would have been 
already 
possible with Suzaku XRS (Kelley et al.\ 2007), and will be certainly
realized by future calorimeter missions such as NeXT (Inoue and Kunieda 2004).

\section{Summary}
We have observed the typical Galactic plane region at $(l,b)\approx(\timeform{28.46D},\timeform{-0.20D})$
(the Chandra\/ AO1 deep field)  with Suzaku for 100 ksec to carry out spectral study of the 
Galactic Ridge X-ray Emission (GRXE).  We were able to, for the first time, 
 resolve three narrow iron K-emission lines with different
ionization states from the GRXE, which constrains the GRXE emission models.
Collisional ionization equilibrium plasma is the likely origin of the
He-like and H-like K$\alpha$ lines, while origin of the 6.41 keV line is not elucidated.
Excluding the point sources brighter than $2 \times 10^{-13}$ ergs s$^{-1}$cm$^{-2}$
(2-- 10 keV), total observed flux in the direction of the Galactic plane
(including cosmic X-ray background) 
is $\sim6.1 \times 10^{-11}$ ergs s$^{-1}$ cm$^{-2}$ deg$^{-2}$ (2--10 keV), among which
$\sim4.8 \times 10^{-11}$ ergs s$^{-1}$ cm$^{-2}$ deg$^{-2}$ is considered to be 
Galactic.
%% Point sources brighter than $3 \times 10^{-15}$ ergs s$^{-1}$cm$^{-2}$ (2 -- 10 keV), including
%% both Galactic and extragalactic, can explain only $\sim 15 \%$ of the total observed flux
%% on the Galactic plane. 
% Integrating  the known Galactic point source populations down to 
% $\sim10^{-17}$ ergs s$^{-1}$ cm$^{-2}$ (2 -- 10 keV)  may not account for all the
% observed GRXE flux.  
In order to discriminate the point source origin and diffuse origin of the 
GRXE, as an alternative of the Chandra ultra-deep observation,
we propose plasma density diagnostics by resolving He-like iron
line complex, which will be made possible by future X-ray micro-calorimeter missions.

\section{Acknowledgement}
We acknowledge all the Suzaku\/ team members for having made 
 this observation and study available.
In particular, we thank Dr.\   Greg Brown and Prof.\ Manabu Ishida
for useful comments on plasma diagnostics.


\begin{thebibliography}{}
%% \bibitem[Anders, Grevesse(1989)]{Anders1989}   Anders, E., \& Grevesse, N.\ 1989, Geochim. Cosmochim. Acta, 53, 197
%% \bibitem[Asai et al.(2000)]{Asai2000}   Asai, K., Dotani, T., Nagase, F., \& Mitsuda, K.\ 2000, \apjs, 131, 571
\bibitem[Bamba et al.\ 2001]{Bamba2001}Bamba, A., Ueno, M., Koyama, K., \& Yamauchi, S.\ 2001, \pasj, 53, L21
%% \bibitem[Dame et al.(2001)]{Dame2001}   Dame, T. M., Hartmann, D., \& Thaddeus, P.\ 2001, \apj, 547, 792
%% \bibitem[Dickey, Lockman(1990)]{Dickey1990}   Dickey, J. M., \& Lockmann, F. J.\ 1990, \araa, 28, 215
\bibitem[Ebisawa et al.(2001)]{Ebisawa2001}   Ebisawa, K., Maeda, Y., Kaneda, H., \& Yamauchi, S.\ 2001, Science,    293, 1633
\bibitem[Ebisawa et al.(2005)]{Ebisawa2005}   Ebisawa, K., et al. 2005, \apj, 635, 214
\bibitem[Ezuka, Ishida(1999)]{Ezuka1999}   Ezuka, H., \& Ishida, M.\ 1999, \apjs, 120, 277
%% \bibitem[Favata et al.(2000)]{Favata2000}   Favata, F., Reale, F., Micela, G., Sciortino, S., Maggio, A., \&    Matsumoto, H.\ 2000, \aap, 353, 987
%% \bibitem[G\"udel et al.(1999)]{Gudel1999}   G\"udel, M., Linsky, J. L., Brown, A., \& Nagase, F.\ 1999, \apj, 511, 405
%% \bibitem[Imanishi et al.(2001)]{Imanishi2001}   Imanishi, K., Koyama, K., \& Tsuboi, Y.\ 2001, \apj, 557, 747
%% \bibitem[Ishida(1992)]{Ishida1992}   Ishida, M.\ 1992, Ph. D. thesis, The University of Tokyo 
\bibitem[]{}Fujimoto, R. et al.\ 2007, \pasj, 59, S133
\bibitem[]{}Inoue, H. \& Kunieda, H.\ 2004, Advances Space Research, 34, 2628
\bibitem[]{}Ishisaki Y. et al.\ 2007, PASJ, 59, S113
\bibitem[]{}Kaastra, J. S. \& Mewe, R. \aap,  1993, 97, 443
\bibitem[Kaneda et al.\ 1997]{Kaneda}Kaneda, H., Makishima, K., Yamauchi, S., Koyama, K., Matsuzaki, K., \& Yamasaki, N. Y. 1997, \apj, 491, 638
%% \bibitem[Kaneda et al.(1997)]{Kaneda1997}   Kaneda, H., Makishima, K., Yamauchi, S., Koyama, K., Matsuzaki, K., \&    Yamasaki, N. Y.\ 1997, \apj, 491, 638
%% \bibitem[Kokubun et al.(2006)]{Kokubun2006}   Kokubun, M., et al.\ 2006, \pasj, submitted
\bibitem[Koyama et al.\ 1986]{Koyama}Koyama, K., Makishima, K., Tanaka, Y., \& Tsunemi, H. 1986, \pasj, 38, 121
%% \bibitem[Koyama et al.(1996)]{Koyama1996}   Koyama, K., Hamaguchi, K., Ueno, S., Kobayashi, N., \& Feigelson, E. D.\    1996, \pasj, 48, L87
\bibitem[]{}Kelley, R. L., et al.\ 2007, \pasj, 59,  S77
\bibitem[Koyama et al.(2007a)]{Koyama2007}   Koyama, K., et al.\ 2007a, \pasj, 59, S23
\bibitem[Koyama et al.(2007b)]{Koyama2007}   Koyama, K., et al.\ 2007b, \pasj, 59, S245
%% \bibitem[Mewe et al.(1995)]{Mewe1995}   Mewe, R., Kaastra, J. S., \& Liedahl, D. A.\ 1995, Legacy, 6, 16
\bibitem[Mitsuda et al.(2007)]{Mitsuda2007}   Mitsuda, K., et al.\ 2007, \pasj, 59, S1
%% \bibitem[Morrison, McCammon(1983)]{Morrison1983}   Morrison, R., \& McCammon, D. 1983, ApJ, 270, 119
\bibitem[]{}Moretti, A., Campana, S., Lazzati, D., \& Tagliaferri, G. 2003, \apj, 588, 696
\bibitem[Mukai, Shiokawa(1993)]{Mukai1993}   Mukai, K., \& Shiokawa, K.\ 1993, \apj, 418, 863
\bibitem[Nakajima et al.(2007)]{Nakajima2007} Nakajima, H., et al. 2007, \pasj, this volume
%% \bibitem[Nagase(1989)]{Nagase1989}   Nagase, F.\ 1989, \pasj, 41,1
\bibitem[]{}Ottmann, R. \& Schmitt, J. H. M. M. 1992, \aap, 256, 421
\bibitem[]{}Revnivtsev, M., Sazonov, S., Gilfanov, M., Churazov, E., \& Sunyaev, 
R. 2006 \aap, 452,  169
%% \bibitem[]{Revnivtsev2006-2}Revnivtsev, M.,  Molkov, S. and  Sazonov, S.,      MNRAS, {\bf 373}  2006, L11.
\bibitem[]{RS2007} Revnivtsev, M., \& Sazonov, S. 2007, \aap, 471, 159
%% \bibitem[Revnivtsev et al.(2006)]{Revnivtsev2006}   Revnivtsev, M., Sazonov, S., Gilfanov, M., Churazov, E., \& Sunyaev, R.\    2006, \aap, in press
\bibitem{Sazonov2006}Sazonov, S., Revnivtsev, M., Gilfanov, M., Churazov, E., \& Sunyaev, R. 2006, \aap, 450,  117
%% \bibitem[Skinner et al.(1997)]{Skinner1997}   Skinner, S. L., G\"udel, M., Koyama, K., \& Yamauchi, S.\ 1997, \apj,    486, 886
\bibitem[Serlemitsos et al.(2007)]{XRT}
   Serlemitsos, P., et al.\ 2007, \pasj, 59, S9
%% \bibitem[Stern et al.(1992)]{Stern1992}   Stern, R. A., Uchida, Y., Tsuneta, S., \& Nagase, F.\ 1992, \apj, 400, 321 
\bibitem[]{}Tanaka, Y., Koyama, K., Maeda, Y., \& Sonobe, T. 2000, \pasj, 52, L25
\bibitem[]{Tanaka}Tanaka, Y. 2002, \aap,  382,  1052
\bibitem[Takahashi et al.(2007)]{Takahashi2007}   Takahashi, T., et al.\ 2007, \pasj, 59, S35
\bibitem[]{}Tawa, N., et al.\ 2007, PASJ, this volume
%% \bibitem[Tsuboi et al.(1998)]{Tsuboi1998}   Tsuboi, Y., Koyama, K., Murakami, H., Hayashi, M., Skinner, S., \&    Ueno, S.\ 1998, \apj, 503, 894
%% \bibitem[Tsuru et al.(1989)]{Tsuru1989}  Tsuru, T., et al.\ 1989, \pasj, 41, 679
%% \bibitem[Tanaka, Lewin(1995)]{Tanaka1995}   Tanaka, Y., \& Lewin, W. H. G.\ 1995, in X-Ray Binaries,    ed. W. H. G. Lewin, J. van Pradijs, \& E. P. J. van den Heuvel    (Cambridge: Cambridge Univ. Press), 126
%% \bibitem[]{Ueno}Ueno, M., Bamba, A., Koyama, K. and Ebisawa, K. 2003, \apj, 588, 338
\bibitem[Ueno et al.\ 2003]{Ueno2003} Ueno, M., Bamba, A., Koyama, K., \& Ebisawa, K.\ 2003, \apj, 588, 338
\bibitem[]{}Valinia, A., et al. 2000, \apj, 543, 733
\bibitem[]{}Wargelin, B. J., Beiersdorfer, P., Neill, P. A., Olson, R. E., \& Scofield, J. H. 2005,
\apj, 634, 687
\bibitem[Warwick et al.\ 1985]{Warwick}Warwick, R. S., et al. 1985, Nature,  317,  218.
%\bibitem[White et al.(1986)]{White1986}   White, N. E., Culhane, J. L., Parmar, A. N., Kellett, B. J., Kahn, S.,    van den Oord, G. H. J., \& Kuijpers, J.\ 1986, \apj, 301, 262
\bibitem[Worrall et al.\ 1982]{Worrall}Worrall, D. M., et al. 1982, \apj, 255, 111
\bibitem[]{}Worrall, D. M. \& Marshall, F. E. 1983, \apj, 267, 691
%\bibitem[Yamauchi et al.(1996)]{Yamauchi1996}   Yamauchi, S., Kaneda, H., Koyama, K., Makishima, K., Matsuzaki, K.,    Sonobe, T., Tanaka, Y., \& Yamasaki, N.\ 1996, \pasj, 48, L15
\bibitem[Yamasaki et al.(1997)]{Yamasaki1997}   Yamasaki, N. Y., et al.\ 1997, \apj, 481, 821
\bibitem[]{Yamauchi2007}Yamauchi, S., et al. 2007,  59, S215
\bibitem[]{}Zombeck, M. V. 2007, ``Handbook of Space Astronomy and Astrophysics'', Third Edition,
Cambridge University Press
%% \bibitem{}Yamauchi, K. and Koyama, K.,\AJ{620,1993,404}.
%% \bibitem{}Mukai, K. and Shiokawa, K.,\AJ{418,1993,863}.
%% \bibitem{}Sugizaki, M., Mitsuda, K., Kaneda, H., Matsuzaki, K., Yamauchi, S.
%% \& Koyama, K., Astrophys.\ J.\ Supp. {\bf 134} (2001), 77.
%% \bibitem{}Hands, A. D. P., Warwick, R. S., Watson, M. G. and Heldfand,D. J., 
%% MNRAS {\bf 351} (2004), 31\\
%% \bibitem{}Moretti, A., Campana, S., Lazzati, D. and Tagliaferri, G.,
%% \AJ{588,2003,696}.
%% \bibitem{}Mukai, K. and Shiokawa, K., \AJ{418,1993,863}
%\bibitem[Bailey(1981)]{Bailey1981}
%   Bailey, J.\ 1981, \mnras, 197, 31
%\bibitem[Beuermann, Weichhold(1999)]{Beuermann1999}
%   Beuermann, K., \& Weichhold, M.\ 1999, in Cataclysmic Variables, 
%   ed. C. Hellier, \& K. Mukai, ASP conference Series 157, 283
%\bibitem[Henry(1991)]{Henry1991}
%   Henry, T. J.\ 1991, Ph. D. thesis, Univ. Arizona
%\bibitem[Henry, McCarthy(1990)]{Henry1990}
%   Henry, T. J., \& McCarthy, D. W., Jr.\ 1990, \apj, 350, 334
%\bibitem[Ishida et al.(1997)]{Ishida1997}
%   Ishida, M., Matsuzaki, K., Fujimoto, R., Mukai, K., \& Osborne, J. P.\ 
%   1997, \mnras, 287, 651
%\bibitem[Koyama et al.(1986)]{Koyama1986}
%   Koyama, K., Makishima, K., Tanaka, Y., \& Tsunemi, H.\ 1986, \pasj, 38, 121
%\bibitem[Leggett(1992)]{Leggett1992}
%   Leggett, S. K.\ 1992, \apjs, 82, 351
%\bibitem[Mukai et al.(1994)]{Mukai1994}
%   Mukai, K., Ishida, M., \& Osborne, J. P.\ 1994, \pasj, 46, L87
%\bibitem[Osborne et al.(1994)]{Osborne1994}
%   Osborne, J. P., Beardmore, A. P., Wheatley, P. J., Hakala, P., 
%   Watson, M. G., Mason, K. O., Hassall, B. J. M., \& King, A. R.\ 1994, 
%   \mnras, 270, 650
%\bibitem[Ramseyer(1994)]{Ramseyer1994}
%   Ramseyer, T. F.\ 1994, \apj, 425, 243
%\bibitem[Sakano et al.(2002)]{Sakano2002}
%   Sakano, M. et al.\ 2002, \apjs, 138, 19
%\bibitem[Sparks et al.(1999)]{Sparks1999}
%   Sparks, W. M., Starrfield, S. G., Sion, E. M., Shore, S. N., Chanmugam, G., 
%   \& Webbink, R. F.\ 1999, Allen's Astrophysical Quantities Fourth Edission 
%    (ed. A. N. Cox), AIP Press, p.429 
%\bibitem[Sugizaki et al.(2001)]{Sugizaki2001}
%   Sugizaki, M. et al.\ 2001, \apjs, 134, 77
%\bibitem[Strom et al.(1995)]{Strom1995}
%   Strom, K., M., Kepner, J., \& Strom, S. E.\ 1995, \apj, 438, 813
%\bibitem[Valinia, Marshall(1998)]{Valinia1998}
%   Valinia, A., \& Marshall, F. E.\ 1998, \apj, 505, 134
%\bibitem[Worrall et al.(1982)]{Worrall1982}
%   Worrall, D. M., Marshall, F. E., Boldt, E. A., \& Swank, J. H.\ 1982, \apj
%   255, 111
%\bibitem[Yamauchi, Koyama(1993)]{Yamauchi1993}
%   Yamauchi, S., \& Koyama, K.\ 1993, \apj, 404, 620
\end{thebibliography}
\end{document}